# MAGNETIC PROPERTIES OF $Fe_{75}Cu_4Ni_4Mo_2B_{15}$ ALLOY

B. Poornachandra Sekhar and G. Markandeyulu*
*Magnetism and Magnetic Materials Laboratory*
*Department of Physics*
*Indian Institute of Technology Madras*
*CHENNAI – 600 036, India*

## Abstract

The structural and magnetic properties of ball-milled $Fe_{75}Cu_4Ni_4Mo_2B_{15}$ alloy have been investigated through powder X-ray diffraction studies, magnetization and magnetoimpedance measurements. The particle size decreased rapidly with ball-milling up to 165 hours and then remained as a constant with further hours of milling while, the lattice expanded with milling time up to 165 hours and contracted back with further milling. Saturation magnetization is found to decrease rapidly up to 165 hours of milling and then started decreasing slowly with further milling time. Magnetoimpedance studies were carried out in pellets and thin films of $Fe_{75}Cu_4Ni_4Mo_2B_{15}$. Magnetoimpedance in pellet is observed to be ~2.5% at 5Hz whereas, in thin films, it is observed to be ~1.5% at 3 MHz.

* Corresponding author: Tel: +91 – 44- 22578677; FAX: +91 – 44 – 22570509/22570545
E-mail address: mark@physics.iitm.ac.in

## INTRODUCTION

Soft magnetic materials in the form of wires, ribbons, powders and thin films find various applications based on the ratio of $M_r$ to $M_s$, where $M_r$ is remnant magnetization, $M_s$ is saturation magnetization. Materials with large $M_r/M_s$ are particularly suited to devices such as switch cores, high gain magnetic amplifiers and low frequency inverters. Samples with intermediate $M_r$ /$M_s$ are best applied in high frequency transformers where minimum loses are important. Materials with near zero $M_r/M_s$ values are interesting for applications in filter chokes or loading coils and several electromechanical transducers [1].

One of the properties of interest in soft magnetic materials is the magnetoimpedance. The overall effect of the applied magnetic field is to induce a strong modification in the effective magnetic permeability, a factor that is relevant to determine the field and current distribution within the sample [2]. In a soft magnetic material, the magnetic permeability can change by orders of magnitude due to the application of a small magnetic field, causing considerable variations in the internal fields and current density and consequently, in the impedance of the sample [3]. The magnetoimpedance ratio is given by

$$\frac{\Delta Z}{Z}\% = \frac{Z(H) - Z(0)}{Z(0)} \times 100 \qquad (1)$$

where, Z(H) = Impedance in the presence of a magnetic field and Z(0) = Impedance in the absence of a magnetic field. Giant magnetoimpedance (GMI) in amorphous wires [4,5] ribbons and thin films [6,7] has attracted much attention in the last few years because of its potential applications in magnetic recording heads and sensor elements.

Materials based on Fe-B have been attracting attention for several soft magnetic applications, particularly in amorphous form. Several 3d elemental substitutions have been made at the Fe site for modifications in the magnetization, Curie temperature, magnetoimpedance, etc [1]. It has been known that mechanical milling of the powders of crystalline compounds stabilizes the amorphous / nano-crystalline structures through the vacancies and defects created by severe cold work [8]. In this paper, the magnetic properties of ball-milled powders and thin films of $Fe_{75}Cu_4Ni_4Mo_2B_{15}$ are presented.

**EXPERIMENTAL DETAILS**

$Fe_{75}Cu_4Ni_4Mo_2B_{15}$ was prepared by arc melting the constituent elements of high purity (Fe, Mo and B - 99.95%; Cu – 99.9%; Ni – 99.99%). The ingot was melted several times to obtain a homogeneous mixture. The weight loss after melting was less than 0.5%. Structural characterization was carried out by taking powder x-ray diffraction patterns ($MoK_\alpha$ - 0.70930 Å). The computer code Autox was used for indexing the planes.

Ball milling was performed employing a planetary ball mill (FRISTCH) at a speed of 250 rpm and a ball-to-powder weight ratio of 15:1 was used. The milling was carried out in toluene for 310 hours employing tungsten carbide balls of 10 / 20 mm diameter. The magnetization measurements were carried out after different stages of milling hours using a PAR vibrating sample magnetometer up to a field of 10 kOe.

The powder of the as cast alloy was compacted using a hydraulic press, at a force of about 5 tonnes using a stainless steel die. The size of the pellet was 8 mm diameter and 1 mm thick. The thin films were prepared by evaporation to completion technique by taking a small quantity of sample in a spiral boat and evaporating it completely at faster rate. The film dimensions are 20 mm length, 1mm breadth and thickness of about 2500Å. The films were then annealed at $400^0$C in vacuum for one hour. The impedance was measured using a HP 4192A LF IMPEDANCE ANALYSER in the frequency range 5 Hz – 13 MHz.

## RESULTS AND DISCUSSIONS

Fig. 1 shows the powder x-ray diffractograms for $Fe_{75}Cu_4Ni_4Mo_2B_{15}$ alloy. The planes were indexed on the basis of cubic structure. The lattice parameter is found to be 2.60 Å. X-ray diffractograms of the ball-milled samples have been taken at various stages of milling and are shown in Fig. 1. With increasing milling time the intensities of the peaks are seen to decrease and the widths of peaks increase. The average particle sizes have been measured using the Scherrer's formula.

The variation of the lattice parameter with ball-milling is shown in Fig. 2. The value of $\Delta a$ / $a_0$, where $a_0$ is the lattice parameter of the unmilled sample and $\Delta a = a_0 - a_t$ ($a_t$ is the lattice parameter of the sample milled for a time t), decreases gradually with milling time up to 165 hours and then starts decreasing with further milling. The initial increase in lattice parameter with milling time is due to the disordering of the alloy as observed by Olezak [9] in Fe-Al and Chinnasamy *et al* in $Ni_3Fe$ alloys [10]. The variation of lattice parameter in solid solutions composed of atoms with different atomic radii has been explained on the basis of strain relaxation theory by Cahn [11]. According to this theory, during the initial stages of milling, the strains involved in the sample may be large, causing an increase in the lattice parameter. Further milling relaxes the strain which causes a progressive shortening of the bonds and hence an effective decrease of the lattice parameter [12].

Fig. 3 shows the variation of particle size with milling time. It is found that the particle size decreases rapidly with increasing milling time up to 165 hours and then it remains constant with further hours of milling. A similar behavior in particle size has been reported by Chinnasamy *et al.* in $Ni_3Fe$ alloy [13].

The room temperature magnetization curves for the ball milled $Fe_{75}Cu_4Ni_4Mo_2B_{15}$ are shown in Fig. 4. The magnetization of un-milled as well as in the milled samples do not completely saturate up to 1 T. However, the magnetization values (at 1 T) of the samples milled for more than 250 hours are more than those for the other samples. The variation of magnetization with milling times is shown in Fig. 5. It is seen from the figure that the magnetization decreases from a value of 165 emu/g (for un-milled) to 73 emu/g (for 310 hours) with increasing milling time.

The reduction in magnetization of ball milled samples may be attributed to the deviation of interatomic spacing (due to the lattice expansion and subsequent contraction) in the interfacial regions as compared to the crystalline component. This results in the decrease of the long-range exchange interaction. The magnetism in this material being itinerant due to the presence of Fe and Ni, the magnetic moment depends on the exchange interactions that depend on the interatomic distances. Therefore, in the present case, the magnetization is expected to decrease with milling time. Moreover, the reduction in particle size causes disordering at the surfaces and the bulk of the particle due to milling which causes a decrease in magnetization. Another factor is the possible oxidation of the sample during milling, which could not be identified by the x-ray analysis. Although the mechanical milling was performed in toluene environment some degree of oxidation is inevitable for extremely small particles which may cause appreciable decrease in $M_s$ Value [14].

Figs. 6 and 7 show the variation of impedance (Z) of the pellet and the thin film of the above alloy, with frequency (ù). It is seen that the impedance increases with frequency, reaches a maximum and then tends to decrease at higher frequencies (MHz region).

The impedance depends on the frequency of alternating current and transverse magnetic permeability developed due to the applied alternating current along the length of the sample as

$$Z \,\alpha \sqrt{\omega\mu_{\phi}} \qquad (2)$$

In addition, the skin depth known to depend on ù and $\mu_t$ as

$$\delta = \sqrt{\frac{2}{\sigma\mu_{\phi}\omega}} \qquad (3)$$

In the case of a ferromagnetic material, the transverse magnetic field due to the applied current, h is given by [3]

$$h = \frac{j(r)t}{2} \qquad (4)$$

where, j(r) is the current density at a distance r with in the sample and t is the thickness of the conductor. Here the current density j(r) is not uniform throughout the thickness of the conductor and it depends on the frequency of the applied current. Therefore, in the low frequency regime, skin depth will be large and hence the current distribution will be throughout the cross-sectional area of the sample, which causes the effective impedance to be low. In the high frequency regime, the lower skin depth causes the current distribution to be limited only to the surface of the sample and hence the impedance is low.

Figs.8 and 9 show the variation of magnetoimpedance ratio (MIR) for the pellet and the thin film with the external magnetic field. In the case of pellet the maximum MIR is found to be 2.4 % at 5 Hz and at all other frequencies the MIR is seen not to vary appreciably whereas in the case of thin film the maximum MIR is found to be 1.4% at 3 MHz. Constancy of MIR may be due to a very loose coupling between the grains probably due to the surface oxidation in the grains.

In both the cases we have observed the maximum impedance ratio is positive. The demagnetizing fields in a homogeneous sample will force the easy direction of magnetization (EMD) to lie in the plane of the sample. In the unlikely case where the easy direction of magnetization (EMD) is perpendicular to the plane of the sample, the applied DC field, in the plane of the sample in the present case, will cause the rotation of the magnetic moments towards the field direction, increasing the circular permeability and therefore the impedance. Therefore, in the present case, the sample may be inhomogeneous by way of either composition or the structural distribution.

**CONCLUSIONS**

The lattice of $Fe_{75}Cu_4Ni_4Mo_2B_{15}$ increases rapidly with time at the initial stages of milling due to disordering of the alloy and further milling leads to the decrease of the lattice parameter which may be due to the relaxation of the strains involved in the sample. The particle size is found to decrease rapidly with initial milling time and found to be constant with further milling time. The decrease in the magnetization with milling time is attributed to the decrease in the long range exchange interaction with the reduction in particle size as well as the changes in the lattice parameter. Magnetoimpedance has been investigated in thin films and pellets of $Fe_{75}Cu_4Ni_4Mo_2B_{15}$. A maximum positive MIR ratio of 2.4 % at 5 Hz is observed in the case of pellet and 1.4 % at 3 MHz in the case of thin films.

## ACKNOWLEDGEMENT

The authors thank the Department of Physics, Indian Institute of Technology Madras for providing the experimental facilities.

## FIGURE CAPTIONS

1. XRD patterns of ball-milled $Fe_{75}Cu_4Ni_4Mo_2B_{15}$.
2. Variation of lattice expansion with milling time.
3. Variation of particle size with ball-milling time.
4. Magnetization curves of ball-milled $Fe_{75}Cu_4Ni_4Mo_2B_{15}$.
5. Variation of magnetization with milling time.
6. Variation of impedance with frequency for $Fe_{75}Cu_4Ni_4Mo_2B_{15}$ pellet, curve in the inset show the variation in the frequency range 0.01-0.3 kHz.
7. Variation of impedance with frequency for $Fe_{75}Cu_4Ni_4Mo_2B_{15}$ thin film, curve in the inset show the variation in the frequency range 5-45 kHz.
8. Magnetoimpedance ratio of annealed $Fe_{75}Cu_4Ni_4Mo_2B_{15}$ pellet.
9. Magnetoimpedance ratio of $Fe_{75}Cu_4Ni_4Mo_2B_{15}$ thin film.

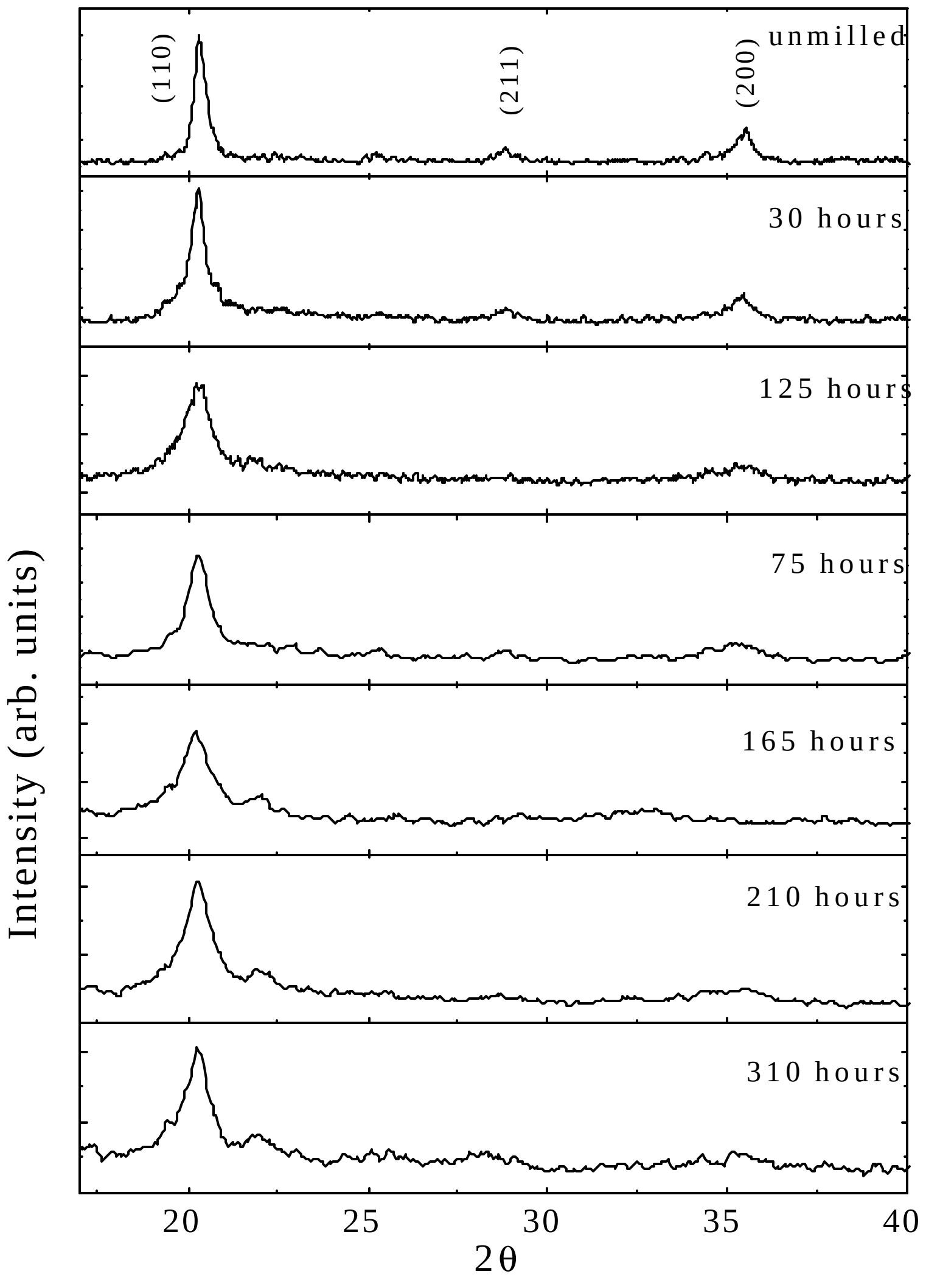

Fig. 1

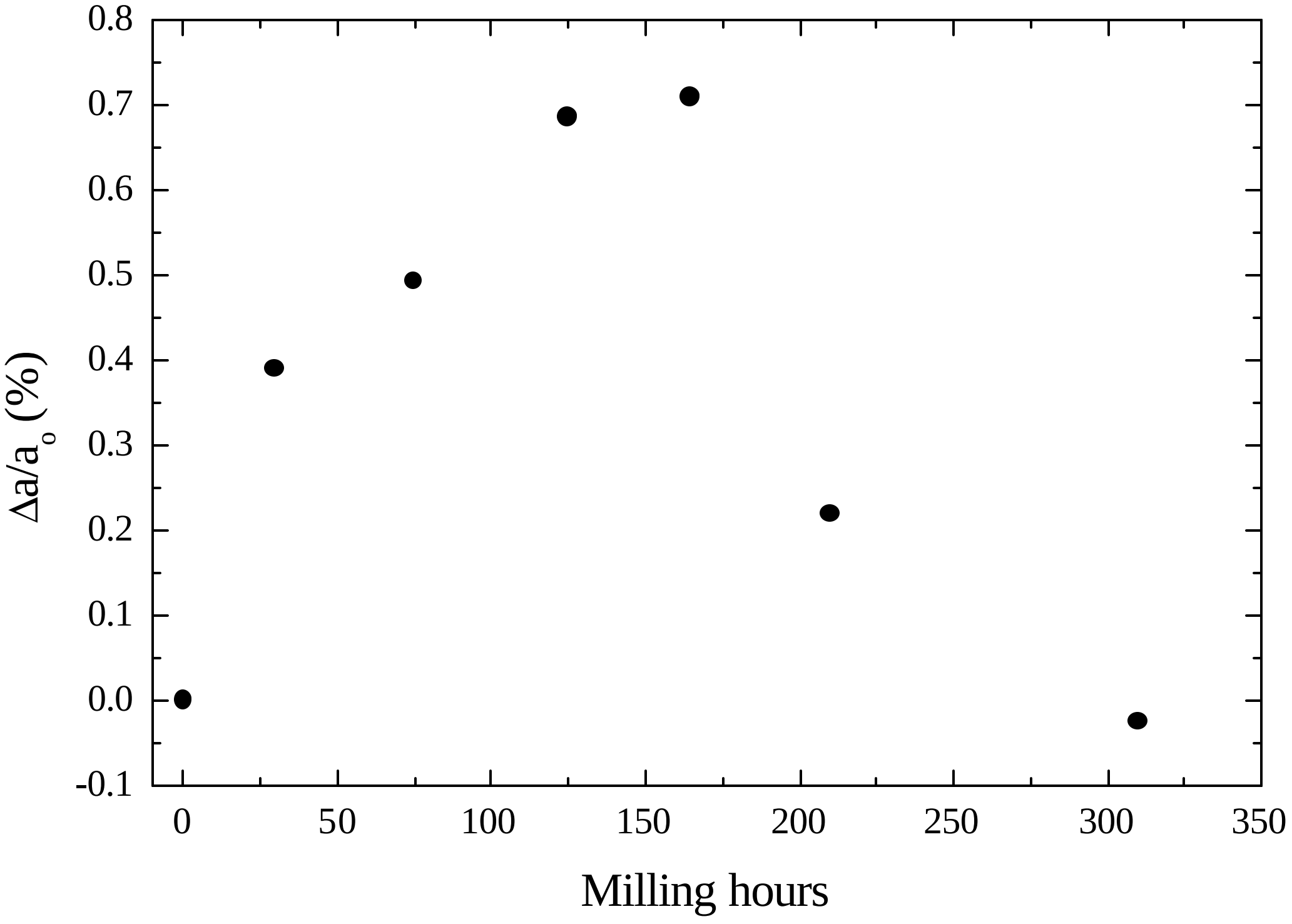

Fig. 2

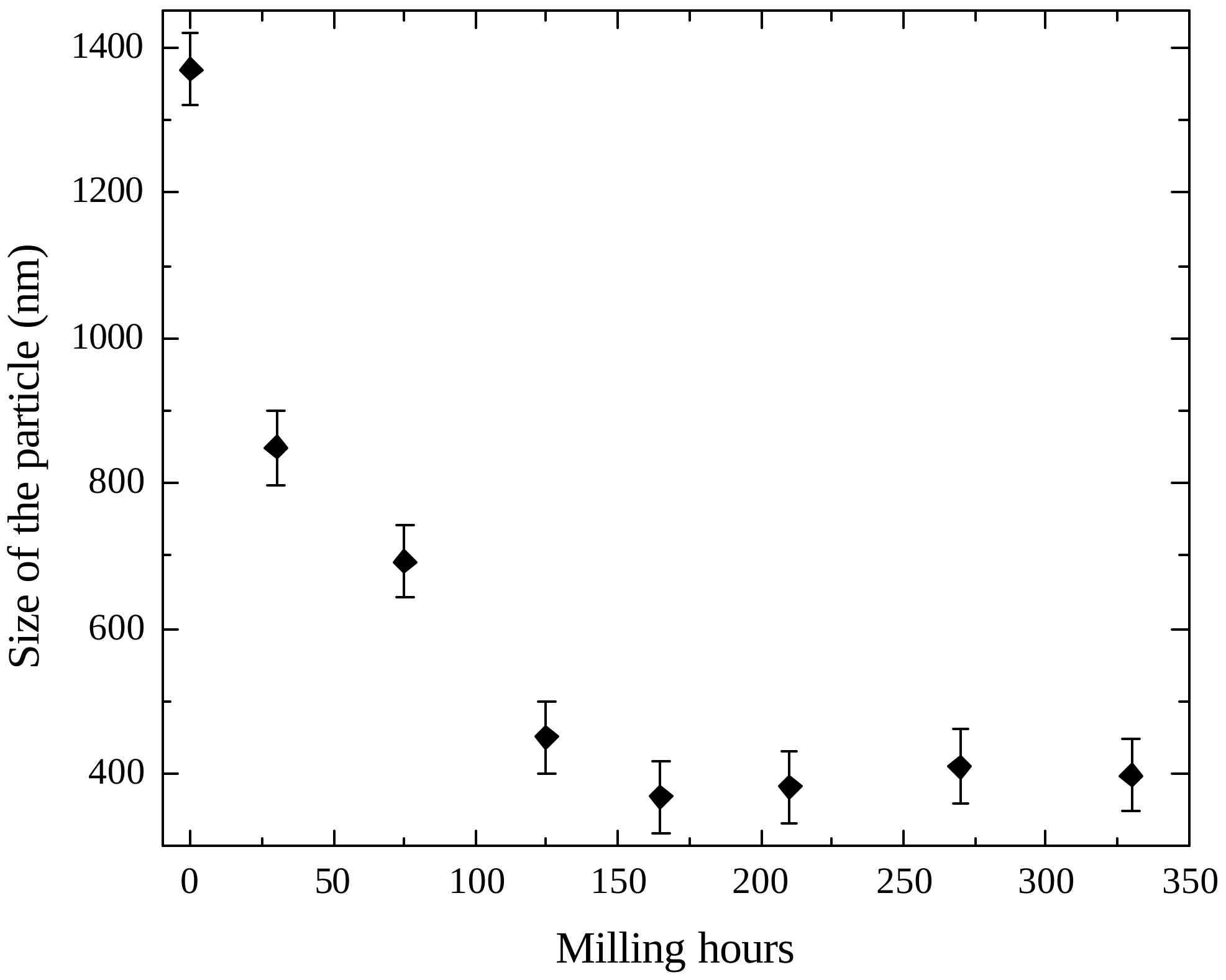

Fig.3

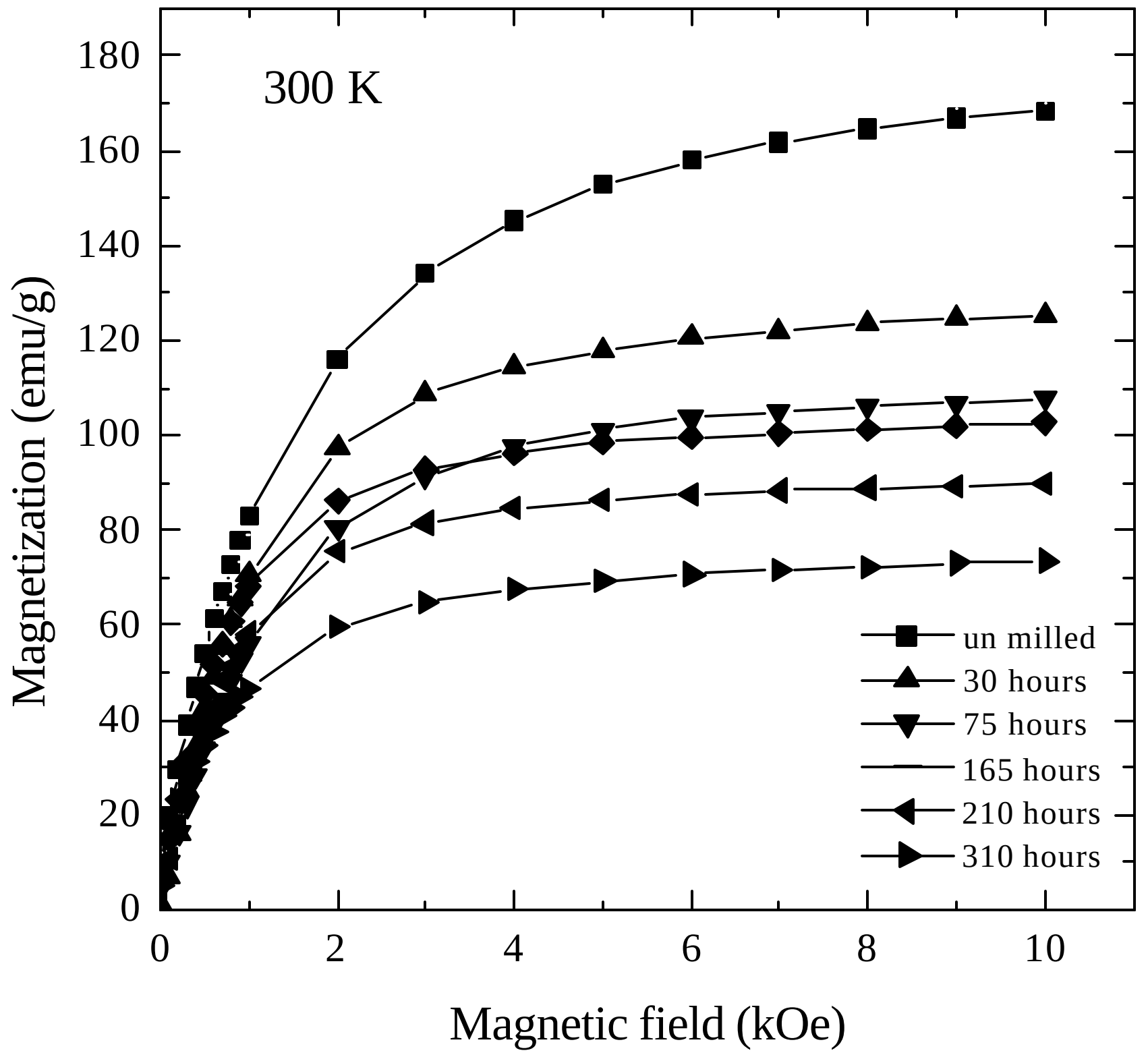

Fig. 4

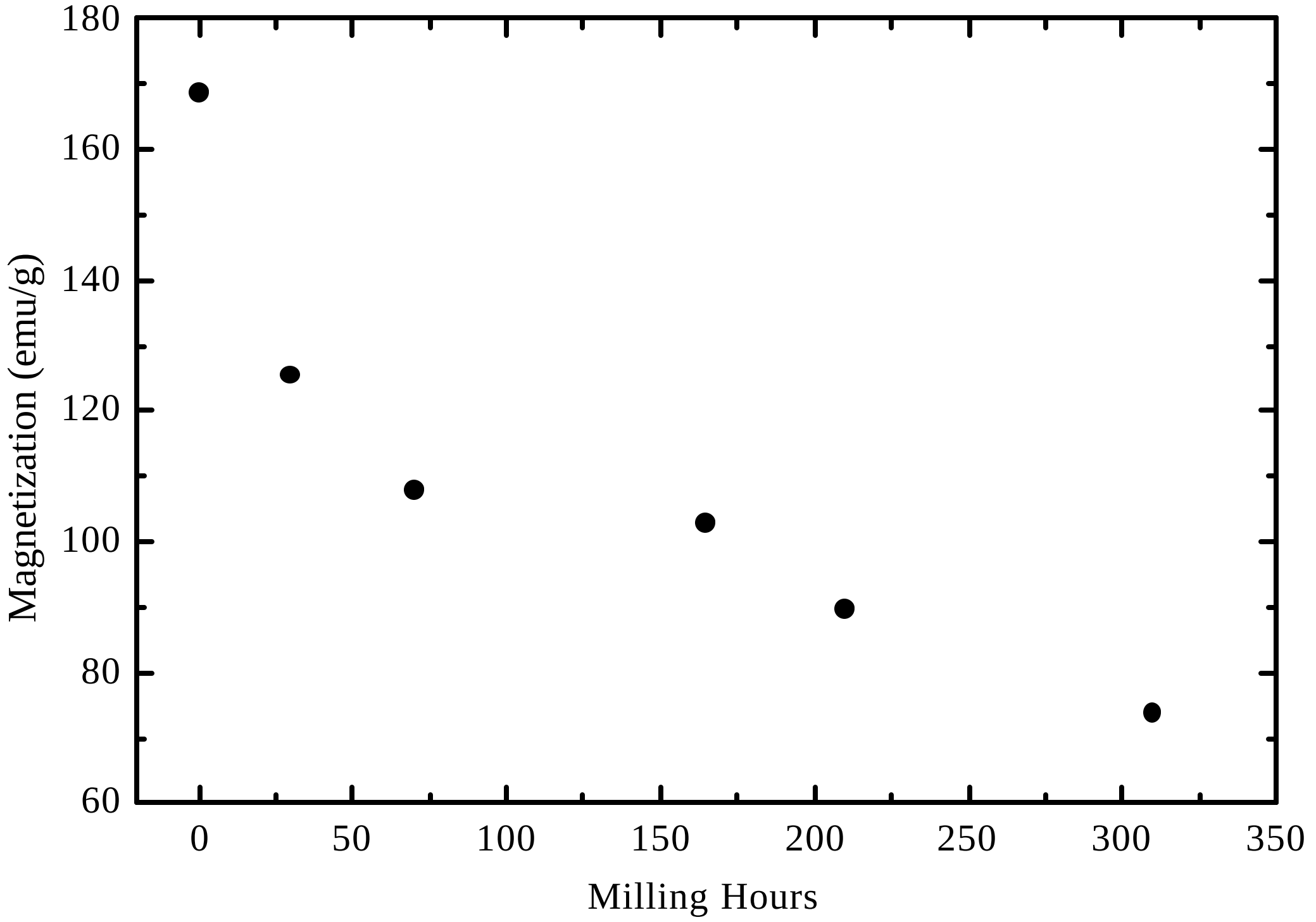

Fig. 5

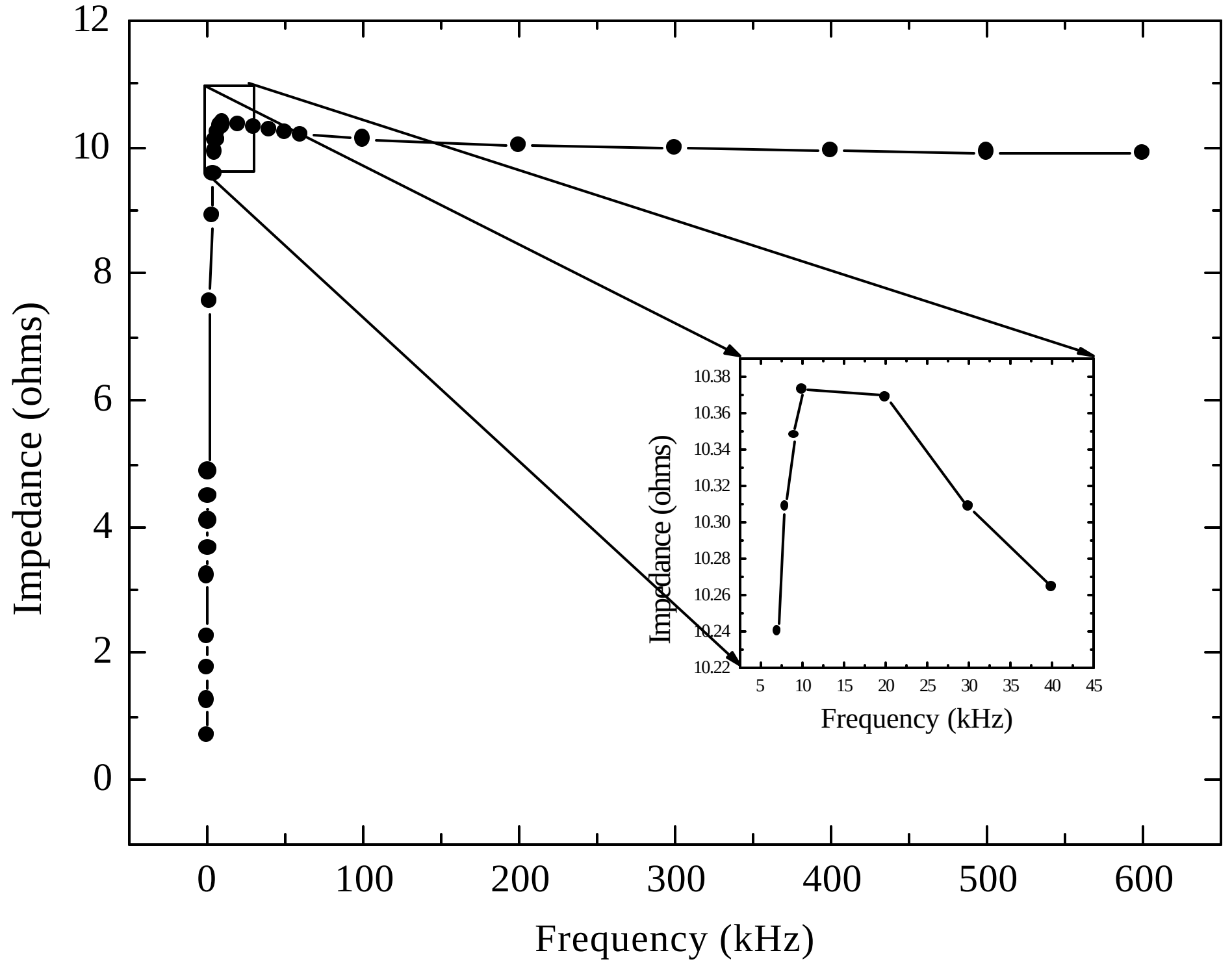
Impedance (ohms)
Frequency (kHz)
Impedance (ohms)
Frequency (kHz)

Fig. 6

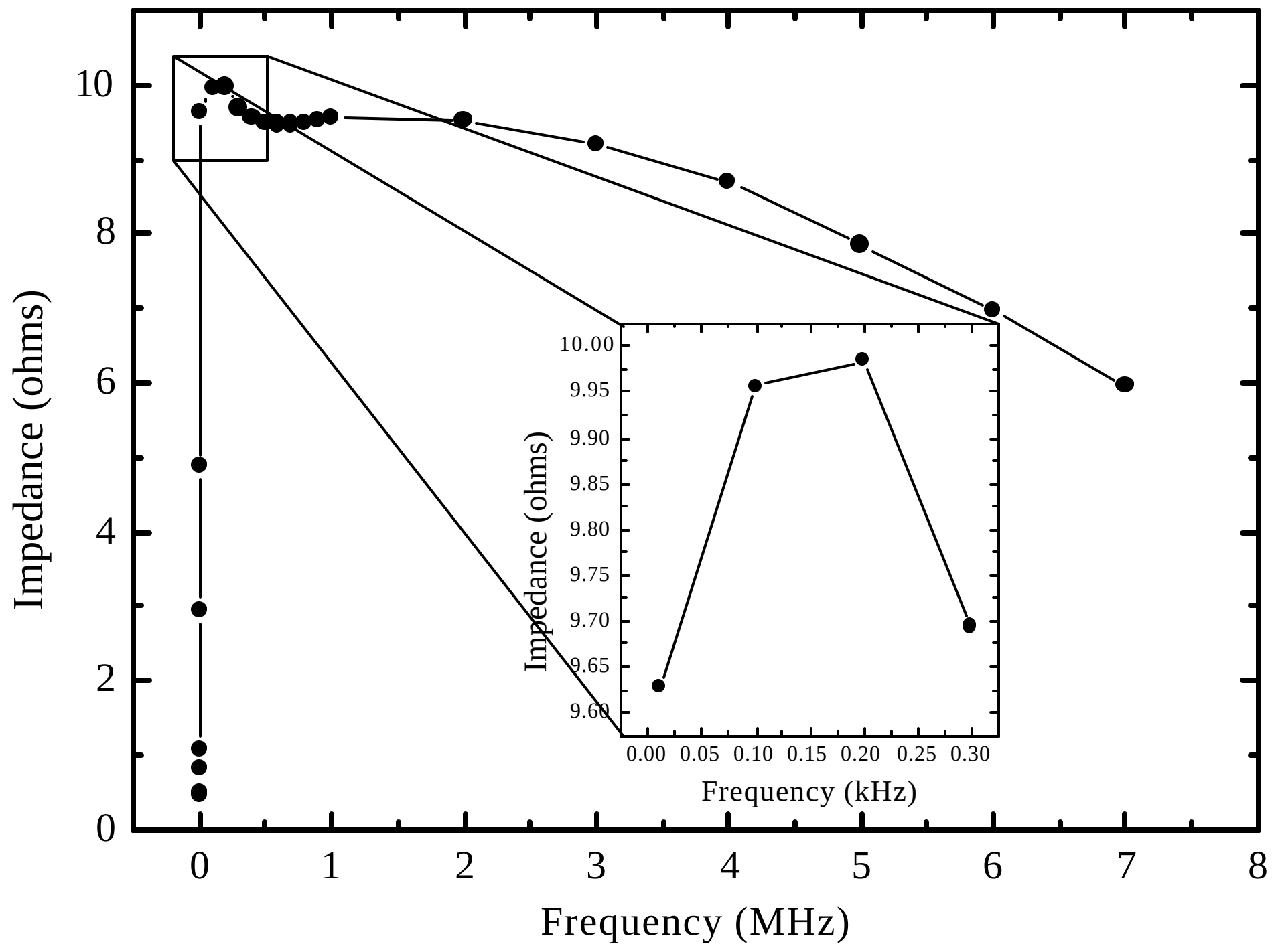
Impedance (ohms)
Frequency (MHz)
Impedance (ohms)
Frequency (kHz)

Fig. 7

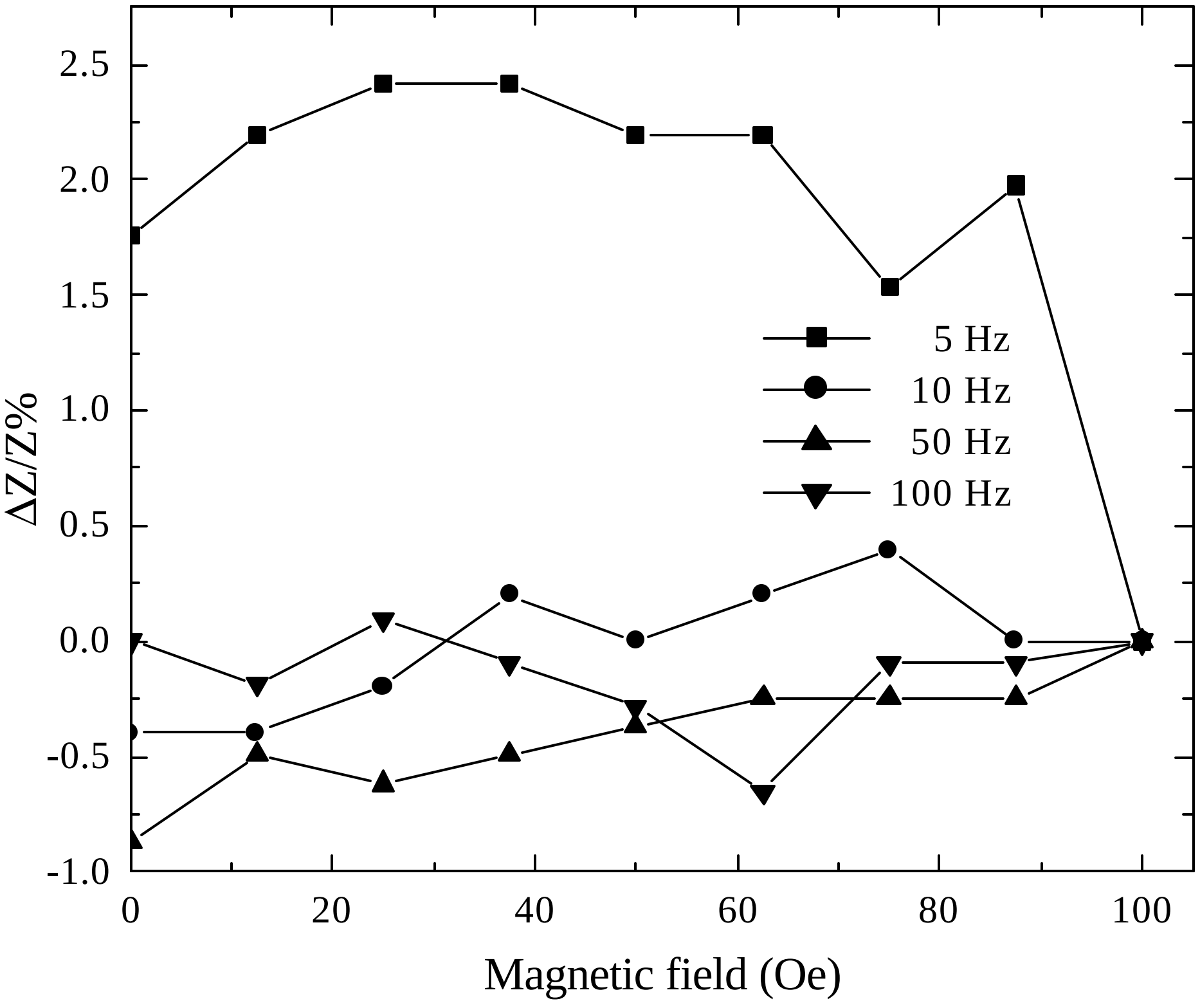

Fig.8

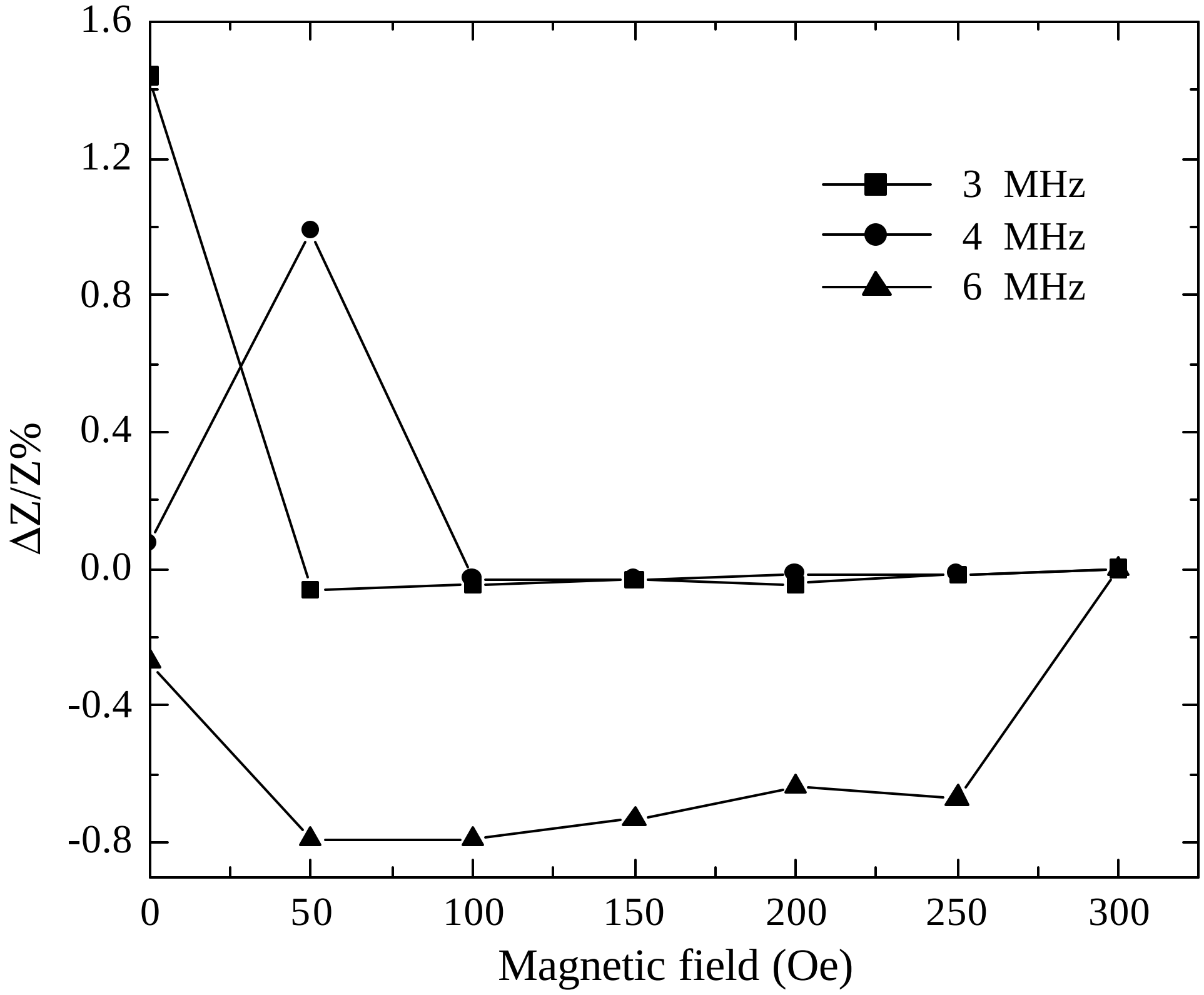

Fig. 9